\documentstyle[12pt,epsfig]{article}

\def\be{\begin{equation}}
\def\ee{\end{equation}}
\def\bq{\begin{eqnarray}}
\def\eq{\end{eqnarray}}
\def\ra{\rightarrow}
\def\n{\nonumber}
\def\g{\gamma}
\def\vp{\varphi}
\def\hx{\hat{x}}
\def\hz{\hat{z}}

\begin{document}

\begin{flushright}
SPhT-97/010 \\
hep-ph/9702207\\
January 1997
\end{flushright}
\vspace{1cm}
\begin{center}
{\bf Distribution of valence quarks and light-cone QCD sum rules. }
\end{center}
\begin{center}{V.M. Belyaev$^*$ }
\end{center}
 \begin{center}
 {\em SPhT, CEA-SACLAY, 91191 Gif-sur-Yvette, CEDEX, France}
\end{center}
\begin{center}{Mikkel B. Johnson}  \\
{Los Alamos National Laboratory, Los Alamos, NM 87545, USA
}
\end{center}
\vspace{1cm}
\begin{abstract}
A method for calculating the pion structure function directly in terms of 
light-cone wave functions is suggested.  Taking twist-2 and twist-4 pion
light-cone wave functions into account, it is shown that the QCD sum rule
prediction is in agreement with quark distribution  obtained from analysis of the Drell-Yan process. Twist-4 quark-gluon light-cone wave 
functions give a large positive contribution to the pion structure function for $x_B<0.2$. 
A new constraint on the twist-2 pion light-cone wave function is obtained. 
We argue that the leading twist pion  wave function 
$\vp_\pi(u)\simeq 1$ for $u=0.3$ with an accuracy of about 20-30\%.

\vspace{0.5cm}

\noindent PACS number(s): 11.15.Tk, 11.55.Hx,  12.38.Lg, 13.60.Hb
\end{abstract}

\vspace{1cm}
\flushbottom{$^*\overline{On\;  leave\;  of\;  absence }
\;from\;  ITEP,\;  117259\;  Moscow,\; Russia.$}

 \newpage

\section{Introduction}
The calculation of structure functions for deep-inelastic lepton-hadron 
scattering,
which is equivalent to finding the quark and gluon distributions in hadrons,
is one of the most important problems of quantum chromodynamics (QCD).
Perturbative treatments of deep-inelastic scattering give important 
information about structure functions as a function on $Q^2$ \cite{evolution} 
($Q^2$ is the momentum squared of the virtual photon) and, under some 
assumptions, can produce reggion behavior at 
small $x_B=Q^2/(2pq)$ \cite{bfkl}.  At the present time there is significant 
activity in 
the application of the perturbative expansion for structure functions \cite{smallx}.  These perturbative investigations are 
very important.  However, since quark and gluon distributions can not be 
determined from perturbative calculations, these distributions 
have been taken from experiment.

Theoretical calculations in QCD have achieved some progress in determining 
the second moments of the quark distribution functions for the 
nucleon \cite{nucl} and pion \cite{pion} from QCD 
sum rules. The first moment of the chiral-odd structure 
function $h_1$ was evaluated in \cite{ji}.   Recently, higher moments of 
quark distribution functions have also been considered in the QCD sum rule 
approach \cite{higher}.

In ref.\cite{ioffe}, Ioffe proposed a four-point correlator for 
theoretical calculations of quark distribution functions in the QCD sum rule 
approach suggested by Shifman, Vainshtein and Zakharov \cite{SVZ}.
This method was applied for calculations of nucleon structure functions, such 
as $F_2(x_B)$ \cite{f2}, $g_1(x_B)$ and $g_2(x_B)$ \cite{g1g2}, 
$h_1(x_B)$ \cite{h1}.  In the present paper we use a light-cone QCD sum rule
\cite{lightcone}. Recently, it was suggested to apply
this version of QCD sum rules to the problem of calculating
the deep-inelastic structure function \cite{bj}.
This  light-cone QCD sum rule is based on the fact (which was 
noted by Ioffe in \cite{ioffe}) that if $x_B$ is not close to the boundary 
$x_B=0$ and $x_B=1$, then the imaginary part of the deep-inelastic 
scattering amplitude is determined  by small distances in the $t$-channel.
In the case of the pion structure function, the nearest singularity in the 
$t$-channel for the correlator of two vector currents and one axial current 
with a pion in the initial state is at $t=-\frac{x_B}{1-x_B}p^2$ for highly 
virtual photons, where $p$ is a momentum of the axial current.
This means that in the case of intermediate $x_B$ we can apply operator 
product expansion (OPE) for the calculation of the correlator in the Euclidean 
region.
Then, we can use a dispersion relation to construct QCD sum rules.
These light-cone QCD sum rules are formulated in terms of so-called
light-cone wave functions of hadrons introduced in perturbative 
QCD to describe hadron form factors at large $Q^2$
\cite{cz,er,bl}.

This new version of QCD sum rules can be a useful tool for investigations 
of quark and gluon distributions and light-cone wave functions in 
hadron physics.  Furthermore, in principle, it is possible to incorporate 
perturbative corrections  to improve the 
calculations.

In the present paper we do not take into consideration perturbative 
corrections.
So, our results can be considered as an initial condition for the
evolution equation \cite{evolution}.
For this reason, at the end of this paper we compare our results with 
the quark distribution function for low $Q^2$ \cite{vogt}
obtained from the the analysis of the Drell-Yan process.

We consider the limit of massless quarks $m_q=0$.  In this limit,
the pion is also massless, $m_\pi=0$.
 
\section{Correlator}

As in \cite{bj}, we consider the correlator
\bq
T_{\mu\rho\lambda}(p,q,k)=
-i\int d^4xd^4ze^{ipx+iqz}
<0|T\{ j_\mu^5(x),j_\rho^d(z),j_\lambda^d(0)|\pi^-(k)>
\label{1}
\eq
for our calculation of the pion structure function.
Here $k$ is the pion momentum, 
\bq
j_\mu^5=\bar{u}\gamma_\mu\gamma_5d,\;\;\;
j_{\rho,\lambda}^d=\bar{d}\gamma_{\rho,\lambda} d ,
\label{2}
\eq
and the following kinematics is used:
\bq
k^2=0; \;\;\; q^2=(p+q-k)^2;\;\;\; t=(p-k)^2=0;\;\;\; s=(p+q)^2;
\n
\\
Q^2=-q^2;\;\;\;(2k,p+q)=s+Q^2; \;\;\; (2pk)=p^2 .
\label{4}
\eq

The discontinuity in $s$ at fixed $p^2$ and $Q^2$ of the correlator (\ref{1}) 
is calculated from
\bq
Im T_{\mu\rho\lambda}=\frac{1}{2i}
\left[
T_{\mu\rho\lambda}(p^2,q^2,s+i\varepsilon)-
T_{\mu\rho\lambda}(p^2,q^2,s-i\varepsilon)
\right] ,
\label{5}
\eq
where $p^2$ and $q^2$ are space-like  vectors, $p^2<0$, $q^2<0$, such that 
$|p^2|,|q^2|\gg \Lambda_{QCD}$.
In the scaling limit, we assume that $|p^2|\ll |q^2|$ and keep only the first 
nonvanishing terms in an expansion in powers of $p^2/q^2$.

$Im T_{\mu\rho\lambda}$ is calculated in the physical region of the 
$s$-channel, and the pion contribution in this amplitude has the following 
form,
\bq
Im T_{\mu\rho\lambda}=p_\mu\frac{f_\pi}{p^2}
Im\left\{
i\int d^4ze^{iqz}<\pi(p)|T\{ j_\rho^d(z),j_\lambda^d(0)\}|\pi(k)>
\right\} .
\label{6}
\eq

On the other hand, the general form for the imaginary part in (\ref{6}) is
\bq
Im\left\{
i\int d^4ze^{iqz}<\pi(p)|T\{ j_\rho^d(z),j_\lambda^d(0)\}|\pi(k)>
\right\}
\n
\\
=A_1(s,Q^2)(q^2g_{\rho\lambda}-q_\rho^{(2)} q^{(1)}_\lambda)
\nonumber
\\
+A_2(s,Q^2)(q^2g_{\rho\lambda}
-q^{(1)}_\rho q^{(1)}_\lambda-q^{(2)}_\rho q^{(2)}_\lambda +
q^{(1)}_\rho q^{(2)}_\lambda)
\nonumber
\\
+
B_1(s,Q^2)\left(p-\frac{pq^{(1)}}{q^2}q^{(1)}\right)_\rho
\left(p-\frac{pq^{(2)}}{q^2}q^{(1)}\right)
_\lambda
\nonumber
\\
+B_2(s,Q^2)\left(p-\frac{pq^{(1)}}{q^2}q^{(1)}\right)_\rho
\left(p-\frac{pq^{(2)}}{q^2}q^{(2)}\right)_\lambda 
\nonumber
\\
+B_3(s,Q^2)\left(p-\frac{pq^{(1)}}{q^2}q^{(2)}\right)_\rho
\left(p-\frac{pq^{(2)}}{q^2}q^{(1)}\right)_\lambda
\nonumber
\\
+B_4(s,Q^2)\left(p-\frac{pq^{(1)}}{q^2}q^{(2)}\right)_\rho
\left(p-\frac{pq^{(2)}}{q^2}q^{(2)}\right)_\lambda,
\label{7}
\eq
where $q^{(1)}=q$ and $q^{(2)}=p+q-k$ are the momenta of the virtual 
photons; $(q^{(1)})^2=(q^{(2)})^2=q^2$.

The structure function of deep-inelastic scattering is defined when 
$q^{(1)}=q^{(2)}$, and it is clear in this case that
\bq
4 x_B^2 q^d(x_B) =\frac{Q^2}\pi\left(B_1(s,Q^2)+B_2(s,Q^2)+B_3(s,Q^2)+B_4(s,Q^2)
\right)_{|Q^2\rightarrow\infty},
\label{8}
\eq
where
\bq
Im\left\{
i\int d^4ze^{iqz}<\pi(p)|T\{ j_\rho^d(z),j_\lambda^d(0)\}|\pi(p)>
\right\}
\nonumber
\\
=
4\pi\frac{x_B^2q^d(x_B)}{Q^2}
\left(p-\frac{pq}{q^2}q\right)_\rho\left(p-\frac{pq}{q^2}q\right)_\lambda
+...~.
\label{9}
\eq
Here $q^d(x_B)$ is $d$-quark distribution function of a pion.

In this paper  the tensor structure $p_\mu p_\rho p_\lambda$ in 
correlator (\ref{1}) is considered.
We define the imaginary part of the correlation function for these tensor 
structures as $f_\pi\frac{4 \pi x_B^2}{Q^2}t(p^2,x_B)$.
Then, the dispersion relation for the function $t$ has the following form,
\bq
t(p^2,x_B)=
\left(\frac{q^d(x_B)}{p^2}+\int  \frac{\rho(s,x_B)}{s-p^2}ds\right) ,
\label{10}
\eq
where the last term  corresponds to the higher-states contribution.

To suppress the contribution of exited states, as usually done in QCD sum 
rules, we will consider instead of $t(p^2,x)$ its Borel transform in $p^2$,
\bq
t(M^2,x_B)=\lim_{Q^2,n\rightarrow\infty,Q^2/n=M^2}
\frac{(-p^2)^{n+1}}{n!}\left(
\frac{d}{dp^2}\right)^n t(p^2,x_B)
\nonumber
\\
=-\left(x_B^2 q^d(x_B)+
\int \rho(s,x_B)e^{-s/M^2}ds\right) .
\label{11}
\eq

The left-hand side  will be calculated in terms of the light-cone pion wave 
function by using the operator product expansion (OPE).
Every new term of the OPE will be suppressed by a factor 
$(\Lambda_{QCD}/M)^{t-2}$, where $t$ is the twist of light-cone wave function.
For the right-hand side, we will use the standard continuum model for higher 
states, whose contribution is suppressed exponentially.  As usual in the QCD 
sum rule approach, we can find the desired physical quantity
(here it will be the quark distribution function) by using a fitting procedure 
in the region for the parameter $M^2$ where the higher states (which are 
estimated in the continuum model) and  higher-twist pion wave function
contributions are small.

\section{Twist-2 light-cone wave function.}

In this section we consider the contribution of the leading twist-2 pion 
light-cone wave function in the QCD sum rule (\ref{11}).

It is clear that in the formal limit when $|p^2|,Q^2\ra\infty$, we can use a 
free $d$-quark propagator.
All interactions with soft nonperturbative gluon fields are suppressed by a 
factor $\Lambda_{QCD}^2/p^2$ or $\Lambda_{QCD}^2/Q^2$.
Perturbative contributions should be taken into account.  However, as pointed 
out in the introduction, we do not take these corrections into consideration, 
assuming that our results can be compared directly to the quark 
distribution function at low $Q^2$, where these contributions are small.

The result of a very simple calculation with free $d$-quark propagators 
in (\ref{1}) gives 
\bq
T_{\mu\rho\lambda}(p,q,k)=i\int d^4xd^4ze^{ipx+iqz}
\frac{(x-z)_\alpha z_\beta}{4\pi^4(x-z)^4z^4}
\n
\\
<0|\bar{u}(x)\g_\mu\g_5
\g_\alpha\g_\rho\g_\beta\g_\lambda d(0)|\pi(k)>.
\label{12}
\eq
According to eqs.(\ref{7},\ref{8}), the interesting contribution corresponds 
to the symmetric part of the amplitude $T_{\mu\rho\lambda}$.
This symmetric part of the product of $\g$-matrices in (\ref{12}) has the 
following form,
\bq
\g_\mu\g_5
\g_\alpha\g_\rho\g_\beta\g_\lambda=
g_{\rho\beta}(g_{\mu\alpha}\g_\lambda+
\n
\\
g_{\lambda\alpha}\g_\mu)
+g_{\beta\lambda}(g_{\mu\alpha}\g_\rho+g_{\rho\alpha}\g_\mu)
\n
\\
+(terms\;\; with\;\;g_{\rho\lambda},g_{\rho\mu},g_{\lambda\mu}).
\label{13}
\eq
Clearly the terms with $g_{\rho\lambda},g_{\rho\mu},g_{\lambda\mu}$ do not 
contribute to the $B_i$ that are defined in eq.(\ref{7}).

From eq.(\ref{13}) it is easy to see that the amplitude (\ref{12}) is 
determined by only one matrix element,
\bq
<0|\bar{u}(x)\gamma_\mu\gamma_5d(0)|\pi(k)>=
ik_\mu f_\pi\int_0^1dve^{-i(kx)v}(\varphi_\pi(v)+x^2g_1(v)+O(x^4))
\nonumber
\\
+f_\pi\left(x_\mu-\frac{x^2}{kx}k_\mu\right)
\int_0^1due^{-i(kx)v}(g_2(v)+O(x^2)) ,
\label{14}
\eq
where $\varphi_\pi(v)$, $g_1(v)$ and $g_2(v)$ are twist-2 and twist-4 
light-cone pion wave functions.
In the formal limit $|p^2|\ra\infty$ the leading contribution is determined 
by the twist-2 pion wave function, $\varphi_\pi(v)$. 
All other terms in the expansion of the amplitude (\ref{14}) are 
suppressed by $(\Lambda_{QCD}^2/p^2)^{t-2}$
($t$ is the twist of a light-cone wave function).

Using the definition (\ref{14}), we obtain the following expression for the 
contribution of the leading twist-2 wave function,
\bq
-f_\pi\int_0^1dv\int d^4xd^4z\frac{\varphi_\pi(v)}{4\pi^4x^4z^4}
e^{i(p-kv)x+i(p+q-kv)z}
\n
\\
\left[
z_\rho(x_\mu k_\lambda+x_\lambda k_\mu)
+z_\lambda(x_\mu k_\rho+x_\rho k_\mu)
\right] .
\label{15}
\eq
Here we have made the shift $x\ra x+z$.
After integration over the coordinates $x$ and $z$ we obtain
\bq
-f_\pi\int\frac{\varphi_\pi(v)dv}{(p-kv)^2(p+q-kv)^2}
\n
\\
\left[
(p+q-kv)_\rho((p-kv)_\mu k_\lambda+
(p-kv)_\lambda k_\mu)\right.
\n
\\
\left.
+(p+q-kv)_\lambda((p-kv)_\mu k_\rho
+(p-kv)_\rho k_\mu))
\right].
\label{16}
\eq
Using the following relations, which hold for the kinematics (\ref{4}),
\bq
(p-kv)^2=p^2(1-v);\;\;\;(p+q-kv)^2=s-(s+Q^2)v,
\label{17}
\eq
we obtain an expression for the coefficient of the tensor 
$p_\mu p_\rho p_\lambda$ in the basis $(p,q^{(1)},q^{(2)})$ (see \ref{7}),
namely,
\bq
-4f_\pi\int_0^1\frac{(1-v)\varphi_\pi(v)}{p^2(s-(s+Q^2)v)}dv.
\label{18}
\eq
From eq.(\ref{18}), it is easy to find the imaginary part of this amplitude,
$t(p^2,x_B)$, as
\bq
\frac{t(p^2,x_B)}{Q^2}=\frac{(1-v)\varphi_\pi(v)}{p^2(s+Q^2)x_B^2}
\n
\\
=\frac{\varphi_\pi(1-x_B)}{Q^2p^2}.
\label{19}
\eq
In eq.(\ref{19}) we have used the relationships
\bq
Im\frac{1}{s-(s+Q^2)v}=-\pi\delta(s-(s+Q^2)v);
\;\;\;
\frac{Q^2}{s+Q^2}=x_B.
\label{20}
\eq

Comparing the result (\ref{19}) with the dispersion relation (\ref{10}),
(and using isotopic symmetry, $\varphi_\pi(v)=\varphi_\pi(1-v)$), one may be
tempted to claim that
\bq
q(x_B)=\varphi_\pi(x_B).
\label{21}
\eq
Relation (\ref{21}) corresponds to a pure  parton picture when the pion 
consists of two free quarks, since
\bq
\int_0^1q(x)dx=1;\;\;\;\int_0^1xq(x)dx=0.5,
\label{22}
\eq
which follow from normalization ($\int_0^1\varphi_\pi(v)dv=1$) 
and the symmetry $\varphi_\pi(v)=\varphi_\pi(1-v)$ of the twist-2 pion wave 
function.
Note, however, the identification (\ref{21}) can not actually be valid,
since the light-cone wave function is not a positive definite function.
From the point of view of the QCD sum rules that are being constructed in this 
paper, (\ref{20}) means that the leading-twist pion wave function determines 
the quark distribution function only in the region where 
$\varphi_\pi(x_B)$ has a positive value.

Note that in the parton model, the quark distribution 
function is equal
to $\vp_\pi^2(x_B)$. Thus, we can expect that the minimal corrections to the 
relation (\ref{21}) (due to the higher-twist light-cone wave functions) will 
occur  in the region where $\vp_\pi(u)\simeq\vp_\pi^2(u)\simeq q(u)\simeq 1$.
From the quark distribution function obtained in \cite{vogt}, one can 
find that $q(x_B)\simeq 1$
for $0.2<x_B<0.3$ with an accuracy of about 20\%.
So, we can expect that the QCD sum rules will be the most reliable in that
region. To check this assumption, we will have to evaluate the contribution
of twist-4 light-cone wave functions.

\section{Twist-4 Wave Functions}

The easiest part of the calculation is the contribution of the two-particle 
twist-4 wave functions $g_1$ and $g_2$ (see (\ref{14})).
The calculations with these wave functions are technically the same 
as those presented in the previous section. 
This correction was calculated in \cite{bj}, and the corresponding
contribution to $t(p^2,x_B)$ has the following form,
\bq
t_4(p^2,x_B)=\frac{4}{p^4}\left(
\frac{g_1(x_B)+G_2(x_B)}{x_B}+\frac12g_2(x_B)-\frac{d g_1(x_B)}{dx_B}
\right)\n
\\=\frac1{p^4}f_4(x_B).
\label{23}
\eq

There are additional corrections corresponding to  three-particle 
twist-4 wave functions.
The general form of these three-particle wave functions can be written 
as follows:  
\bq
<0|\bar{u}(x)g_sG_{\alpha\beta}\g_\g\g_5(z)d(0)|\pi(k)>
\n
\\
=f_\pi(k_\alpha g_{\beta\g}-k_\beta g_{\alpha\g})
\int{\cal D}\alpha_i e^{-ikx\alpha_1-ikz\alpha_3}f(\alpha_i)
\n
\\
+\frac{f_\pi}{(kx)}k_\g(x_\alpha k_\beta-x_\beta k_\alpha)
\int{\cal D}\alpha_i e^{-ikx\alpha_1-ikz\alpha_3}f_x(\alpha_i)
\n
\\        
+\frac{f_\pi}{(kz)}k_\g(z_\alpha k_\beta-z_\beta k_\alpha)
\int{\cal D}\alpha_i e^{-ikx\alpha_1-ikz\alpha_3}f_z(\alpha_i),
\label{24}
\eq
\bq
i<0|\bar{u}(x)g_s\tilde{G}_{\alpha\beta}(z)\g_\g d(0)|\pi(k)>
\n
\\
=-f_\pi(k_\alpha g_{\beta\g}-k_\beta g_{\alpha\g})
\int{\cal D}\alpha_i e^{-ikx\alpha_1-ikz\alpha_3}\tilde{f}(\alpha_i)
\n
\\
-\frac{f_\pi}{(kx)}k_\g(x_\alpha k_\beta-x_\beta k_\alpha)
\int{\cal D}\alpha_i e^{-ikx\alpha_1-ikz\alpha_3}\tilde{f}_x(\alpha_i)
\n
\\        
-\frac{f_\pi}{(kz)}k_\g(z_\alpha k_\beta-z_\beta k_\alpha)
\int{\cal D}\alpha_i e^{-ikx\alpha_1-ikz\alpha_3}\tilde{f}_z(\alpha_i).
\label{25}
\eq
Here and below we have used the notation 
$\int{\cal D}\alpha_i=\int_0^1d\alpha_1d\alpha_2d\alpha_3\delta(1-\alpha_1-
\alpha_2-\alpha_3)$.
Comparing with the standard form for the light-cone wave functions 
(when $z=vx$) that were considered in \cite{bf}, we obtain the following 
relations,
\bq
f(\alpha_i)=\vp_\perp(\alpha_i)\n,
\\
f_x(\alpha_i)+f_z(\alpha_i)=\vp_\parallel(\alpha_i)+\vp_\perp(\alpha_i),
\n
\\
\tilde f(\alpha_i)=\tilde \vp_\perp(\alpha_i)\n,
\\
\tilde f_x(\alpha_i)+\tilde f_z(\alpha_i)=
\tilde \vp_\parallel(\alpha_i)+\tilde \vp_\perp(\alpha_i).
\label{26}
\eq

For practical calculations, it is very convenient to use the following 
formula for the quark propagator in the 
presence of the gluon field \cite{lightcone},
\bq
<0|T\{ q(x)_\alpha^a,\bar{q}_\beta^b(z)\}|0>=
\delta^{ab}\frac{i(\hat{x}-\hat{z})_{\alpha\beta}}{2\pi^2x^4}\n
\\
+
t^{(n)ab}\frac{g}{16\pi^2}\int_0^1duG_{\mu\nu}^n(ux+(1-u)z)
\n
\\
\frac{((1-u)(\hat{x}-\hat{z})\g_\mu\g_\nu+u\g_\mu\g_\nu(\hat{x}
-\hat{z}))_{\alpha\beta}}{(x-z)^2}
\label{26p}
\\
-gt^{(n)ab}\int_0^1 du (x-z)_\mu A_\mu^n(ux+(1-u)z)\frac{\hat{x}-\hat{z}}{2\pi^2(x-z)^4}+
O(g^2).
\label{27}
\eq
The contribution of the term (\ref{26p}) to the diagram where a $d$-quark 
interacts with gluon field between points $x$ and $z$ 
can be written in the following form,
\bq
\frac{1}{32\pi^4}\int_0^1du\int\frac{d^4xd^4z}{(x-z)^2z^4}
e^{ipx+iqz}\n
\\
<0|\bar{u}(x)\g_\mu\g_5G_{\alpha\beta}(ux+(1-u)z)((1-u)(\hx-\hz)\g_\alpha\g_\beta\n
\\
+
u\g_\alpha\g_\beta (\hx-\hz))\g_\rho\hz\g_\lambda d(0)|\pi(k)>.
\label{28}
\eq
The result of our calculations with (\ref{28}) is
\bq
t_g(x_B,p^2)=-\frac{Q^2}{p^4\pi x_B^2}Im\left\{\int\frac{\alpha_2 du{\cal D}\alpha_i
}
{(1-\alpha_1-u\alpha_3)(s-(s+Q^2)(\alpha_1+\alpha_3))}\right.
\n
\\
\left.
[(1-2u)(2\vp_\perp(\alpha_i)+\vp_\parallel(\alpha_i))
-(2\tilde{\vp}_\perp(\alpha_i)+\tilde{\vp}_\parallel(\alpha_i)]
\right\}_{Q^2\ra\infty}
\n
\\
=\frac{-1}{p^4}\int_0^1du\int_0^{1-x_B}\frac{d\alpha_3}
{x_B+(1-u)\alpha_3}
[(1-2u)(2\vp_\perp(\alpha_i)+\vp_\parallel(\alpha_i))
\n
\\
-(2\tilde{\vp}_\perp(\alpha_i)+\tilde{\vp}_\parallel(\alpha_i)]_{
\alpha_2=x_B,\alpha_1=1-\alpha_2-\alpha_3}
\n
\\
=\frac{f_g(x)}{p^4}.
\label{29}
\eq
In this equation we have used the fact that our result depends on the 
functions $f_x$, $f_z$, $\tilde f_x$ and $\tilde f_z$  in the combination 
(\ref{26}) only.

The contribution of the diagram in which a quark interacts with soft gluon 
fields in the 
propagator $<q(z)\bar{q}(0)>$ is suppressed by a factor $1/Q^2$.

The contribution of the last term (\ref{27}) of the quark propagator  is
\bq
\int e^{ipx+iqz}d^4xd^4z
<0|\bar{u}(x)\g_\mu\g_5\frac{\hx-\hz}{2\pi^2(x-z)^4}
\g_\rho
\frac{\hz}{2\pi^2z^4}\g_\lambda
\n
\\
\int_0^1dv(x-z)_\nu A_\nu(xv+(1-v)z)d(0)|\pi(k)>.
\label{29p}
\eq
In the Fock-Schwinger gauge ($x_\mu A_\mu(x)=0$),
the gauge field $A_\nu$ can be expressed in terms
of $G_{\mu\nu}$ as
\bq
A_\nu(xv+(1-v)z)=\int_0^1 udu(vx+(1-v)z)_\delta
G_{\delta\nu}(u(xv+(1-v)z)).
\label{29p1}
\eq
The result of our calculations for (\ref{29p}) is
\bq
t_{g1}(x_B,p^2)=-2\frac{ Q^2}{x_B^2p^4\pi}
Im\left\{\int
\frac{Q^2(udu{\cal D}\alpha_i
\vp_\parallel(\alpha_i))}{(s-(s+Q^2)(\alpha_1+u\alpha_3))^2}\right\}_
{Q^2\ra\infty}.
\label{29p2}
\eq
In eq.(\ref{29p2}) we did not include the part that could not
be represented in terms of the light-cone wave functions $\vp_\parallel$,
 $\vp_\perp$, $\tilde\vp_\parallel$,
 $\tilde\vp_\perp$. This contribution comes from the following integral,
\bq
\frac{f_\pi}{\pi^4}\int u du dv{\cal D}\alpha_i d^4xd^4z
e^{ix(p-k(\alpha_1+uv\alpha_3))+
iz(p+q-k(\alpha_1+u\alpha_3))}
\n
\\
(kz)\frac{x_\mu z_\rho k_\lambda}{x^2z^4}\left(
\frac{f_x(\alpha_i)}{(k,x+z)}+\frac{v f_z(\alpha_i)}{(k,vx+z)}
\right).
\label{29p3}
\eq
Numerical calculations show that the contribution of twist-4 
quark-gluon wave functions
(without the contribution of (\ref{29p3})) cancels the contribution
of the two-particle wave functions in the QCD sum rule for the first
moment of quark distribution function when we  use a self-consitent
set of twist-4 light-cone wave functions (they will be defined in the next Section)
with the parameter $\varepsilon=0$.
It can be proved that there are no higher-twist
corrections to the sum rule for the first moment of the quark distribution
function.
This indicates that the term (\ref{29p3}) should be small, and we can not 
exclude the possibility that its contribution is equal to zero
for this particular set of light-cone wave functions.
In any case, we do not have a model for the wave functions
$f_x$ and $f_z$ that we can use to make realistic numerical estimates for 
the contribution (\ref{29p3}).

The remainder of the contribution (\ref{29p}) has the following
form,
\bq
f_{g1}(x_B)=
\frac{2}{p^4}\int_0^{1-x_B}\frac{d\alpha_3}{\alpha_3}\vp_\parallel(\alpha_i)_{|\alpha_1=\alpha_2-\alpha_3;\alpha_2=x_B}
\n
\\
-\frac{2}{p^4}\int_0^{1-x_B}\frac{d\alpha_3}{\alpha_3^2}\int_{1-x_B-\alpha_3}^{1-x_B}
\varphi_\parallel(\alpha_i){d\alpha_1}_{|\alpha_2=1-\alpha_1-\alpha_3}
\label{29p4}
\eq

\section{Light-Cone QCD Sum Rule.}

Now we can begin our analysis using the QCD sum rule for
the quark distribution function (\ref{11}).
This QCD sum rule has the following form,
\bq
\vp_\pi(x_B)-\frac1{M^2}(f_4(x_B)+f_g(x_B)+f_{g1}(x_B))
=q^d(x_B)+c(x_B)e^{-m_{A_1}^2/M^2}.
\label{30}
\eq
Here we use the results (\ref{19},\ref{23},\ref{29},\ref{29p4}) obtained in 
the previous 
sections.  The last term on the right-hand side of 
(\ref{30}) imitates the higher-states contribution, where
$m$ is the mass of a resonance.
Below, we will assume that $m\sim m_{A_1}=1.25 GeV$.
To find the quark distribution function $q(x_B)$ we have
to analyze the QCD sum rules for all values of $x_B$.

\begin{figure}
\begin{center}
\vspace*{0.0cm}
\epsfig{file=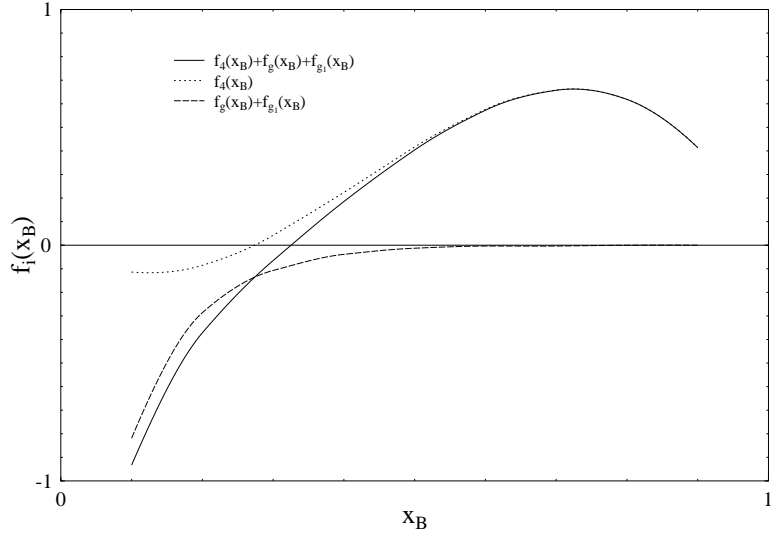,height=9.0cm}
\vspace*{-0.9cm}
\end{center}
\vspace*{-1.5cm}
\caption[]
 {Twist-4 contributions to the d-quark distribution function as
given in Eq. (35), evaluated with the pion wave functions given in Eq. (36).
}
\end{figure}

In Fig.1, the dependence of functions
 $(f_4(x_B)+f_g(x_B)+f_{g1}(x_B))$, $f_4(x_B)$ and 
$(f_g(x_B)+f_{g1}(x_B))$
on $x_B$ is shown.  The QCD sum rule is reliable in the region
$0.2<x_B<0.6$, where the contribution of the twist-4 wave functions is small.
Recall from our earlier discussion that we expected the minimal contribution of
higher-twist wave functions to occur when $0.2<x_B<0.3$.

For our numerical estimates, we have used the following set for light-cone
 wave functions:
\bq
\vp_\pi(u)=6 u(1-u),
\n
\\
g_1(u)=\frac52\delta^2\bar{u}^2u^2
+\frac12\epsilon\delta^2[\bar{u}u(2+13\bar{u}u)
+10 u^3(2-3u+\frac65 u^2)ln(u)\n
\\
+10\bar{u}^3(2-3\bar{u}+\frac65\bar{u}^2)ln(\bar{u})],
\n
\\
g_2(u)=\frac{10}3\delta^2\bar{u}u(u-\bar{u}),
\n
\\
G_2(u)=\frac53\delta^2\bar{u}^2u^2,
\n
\\
\vp_\perp(\alpha_i)=30\delta^2(\alpha_1-\alpha_2)\alpha_3^2
[\frac13+2 \epsilon(1-2\alpha_3)],
\n
\\
\vp_\parallel(\alpha_i)=120\delta^2\epsilon 
(\alpha_1-\alpha_2)\alpha_1\alpha_2\alpha_3,
\n
\\
\tilde{\vp}_\perp(\alpha_i)=30\delta^2\alpha_3^2(1-\alpha_3)
[\frac13+2\epsilon(1-2\alpha_3)],
\n
\\
\tilde{\vp}_\parallel(\alpha_i)=-120\delta^2\alpha_1\alpha_2\alpha_3
[\frac13+\epsilon(1-3\alpha_3)].
\label{31}
\eq
We have used the notation $\bar{u}=(1-u)$.  The wave functions of twist-4 
are very numerous.  Here we use the twist-4 wave functions with leading and 
next-to-leading conformal spin \cite{bf} (see also \cite{gorsky}).

One of the parameters in (\ref{31}) is defined by the matrix element
\bq
<\pi|g_s\bar{d}\tilde{G}_{\alpha\mu}\g_\alpha u|0>=i\delta^2
f_\pi q_\mu.
\label{32}
\eq
The QCD sum rule estimate of ref.\cite{nsvz} yields $\delta^2
=0.2GeV^2$ at $\mu=1 GeV$ ($\mu$ is the renormalization scale).
The last parameter is associated with the deviation of twist-4
wave functions from their asymptotic form. 
Here we used the set with $\epsilon=0$.
It was noted in the previous Section that we can expect that
the contribution of
(\ref{29p3}) is small or even equal to zero.
To use the set with $\epsilon=0.5$ (see Ref. \cite{bf}) we have to
take into account the term (\ref{29p3}).

Note that there is large uncertainty in the choice for the form
of these light-cone wave functions. Even the twist-2 pion
wave function $\vp_\pi(x)$ is not known well enough, and there are different
models for it.  The most peculiar is
the light-cone wave function with a two-humped profile,
which was suggested by Chernyak and Zhitnitsky \cite{cz}. 

The asymptotic form for the wave functions can be taken as 
a starting point for a detailed investigation of these
QCD sum rules. 
\begin{figure}
\begin{center}
\vspace*{0.0cm}
\epsfig{file=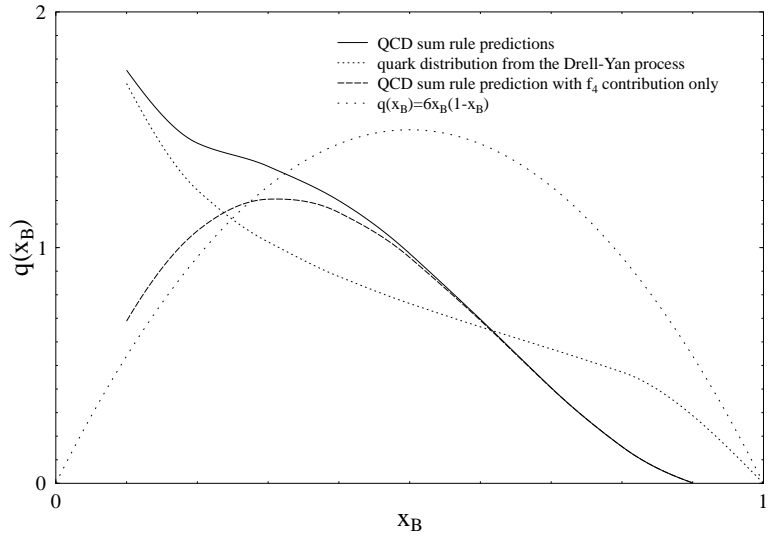,height=9.0cm}
\vspace*{-0.9cm}
\end{center}
\vspace*{-1.5cm}
\caption[]
 {Theoretical results for the d-quark distribution function of the pion
compared to experimental results.  The solid curve is the complete 
result appearing in Eq. (35), evaluated with the wave functions given in 
Eq. (36).}
\end{figure}
The results thus obtained from the sum rule (\ref{30}) are
shown in Fig.2. 
A comparison of these results with the "experimental data",
which were obtained from Drell-Yan process and extrapolated to the low
normalization point \cite{vogt}, shows good agreement.
Agreement with the "experimental data" can be improved by using a
different form for the light-cone wave functions.

It is interesting to estimate the second moment of the quark
distribution functions. Assuming that the region near the end points
$x_B=0,1$ (where our considerations are not valid) gives a small
contribution to the second moment, we 
obtain the following sum rule from eq.(\ref{30}),
\bq
\frac12-\frac{\delta^2}{M^2}+O(M^{-4})=M_2^d+(higher\;\;  resonances),
\label{mom}
\eq
where
\bq
M_2^d=\int_0^1 x_B q^{d}(x_B)dx_B.
\label{d}
\eq
Note that the contribution of the quark-gluon wave function is small.
It is about 10\% of the contribution of twist-4 two particle wave functions.
This means that our QCD sum rule is almost the same as the one obtained in \cite{bj}.
The numerical evaluation of this sum rule gives $M_2^d=0.27\pm 0.05$,
which is in good agreement with "experimental data": $M_2^d\simeq 0.3$ (see
\cite{vogt}).

It was noted above that from the point of view of the partonic model
we can expect minimal higher-twist corrections to occur in the
region where $q(x_B)\simeq 1$. The "experimental data" presented in Fig.2
show that for $x_B\simeq 0.3$, $q(x_B)=1$ .
 The twist-4 correction is small for $x_B=0.3$.  Thus, we can conclude that 
our consideration is self-consistent and leads to
a new constraint on the pion twist-2 light-cone wave function, namely, 
\bq
\vp_\pi(u)\simeq 1 \;\;\;(u=0.3).
\label{const}
\eq
This constraint follows from the "experimental data" \cite{vogt} and from
the sum rule (\ref{30}).
The estimate of higher-twist effects confirms our earlier argument in favor
of such a constraint.

At the present time there are three different models for the twist-2 pion 
wave functions.  The first is asymptotic wave function,
\bq
\vp^{as.}_\pi(u)=6 u(1-u)\;\;\;\vp_\pi^{as.}(0.3)=1.26.
\label{as}
\eq
The second is Chernyak-Zhitnitsky wave function \cite{cz},
\bq
\vp_\pi^{CZ}(u)=30u(1-u)(1-2u)^2;\vp_\pi^{CZ}(0.3)=1.01.
\label{cz}
\eq
And, the third was suggested by Braun and Filyanov \cite{bf},
\bq
\vp_\pi^{BF}(x)=6x(1-x)(1+a_2(\mu) 3/2(5(2 x-1)^2-1)\n
\\
+a_4(\mu) 15/8(21(2 x-1)^4-14(2 x-1)^2+1))
\n
\\
\vp_\pi^{BF}(0.3)=0.72
\label{bf}
\eq
The coefficients $a_2\simeq 0.41$, $a_4=0.23$
correspond to the normalization point $\mu=1.3GeV$ (see \cite{bbkr}).

According to our 
findings, the most preferable model 
for the pion wave function is asymptotic one (\ref{as}).

\section{Conclusion}

In this paper we have presented a new version of the light-cone
QCD sum rule that can be used for investigating the structure
functions of deep-inelastic scattering.
We demonstrated that this approach gives predictions for the quark
distribution function that are in reasonable agreement with 
present "experimental data".

We noted that the present QCD sum rule provides the possibility
to determine the numerical value of the light-cone wave functions
in the region $0.2<x<0.6$, where it is expected that the QCD sum rule should 
work.  However, we have to emphasize that the perturbative corrections may 
give a significant contribution.
These perturbative corrections can be evaluated
in formalism of string operators \cite{lepage}.

Our analysis using the sum rule (\ref{30}) can give
important information about pion wave functions
that are currently widely used in calculations of
different hadronic processes \cite{all}.
A new constraint for the twist-2 light cone wave functions was formulated
from this sum rule.  
This constraint indicates that the twist-2 pion light-cone wave function
is not very different from its asymptotic form.

The agreement with "experimental data" can be significantly improved by changing
the parameter set for the the light-cone wave functions.
It is important to evaluate perturbative corrections.

\section{Acknowledgements}
One of the authors (V.B.) thanks R.Peschanski for useful
comments.
This research was sponsored in part by the U.S. Department of Energy
at Los Alamos National Laboratory under contract 
W-7405-ENG-36.


\begin{thebibliography}{99}


\bibitem{evolution} G. Altarelli, Phys.Rep. {\bf 81} (1982) 1.
\bibitem{bfkl} Ya.Ya. Balitskii and L.N. Lipatov,
Sov.J.Nucl.Phys. {\bf 28} (1978) 822.
\bibitem{smallx} Yu.L. Dokshitzer, V.A. Khoze, A.H. Mueller,
S.I. Troyan, book:
{\it Basics of Perturbative QCD} (Editions FRONTIERES, France);
L.N. Lipatov, preprint DESY-96-132, E-Print Archive: hep-ph/9610276.
\bibitem{nucl} V.M. Belyaev and B.Yu. Blok, Z.Phys. {\bf C30} (1986) 279.
\bibitem{pion} V.M. Belyaev and B.Yu. Blok, Phys.Lett. {\bf B167} (1986) 99.
\bibitem{ji} Hanxim He and Xiangdong Ji,
Phys.Rev {\bf D52} (1995) 2960; Phys. Rev. {\bf D54} (1996) 6897.
\bibitem{higher} G.G. Ross and N. Chamoun, Phys. Lett. {\bf B380} (1996) 151;
N. Chamoun, E-Print Archive: hep-ph/9612467 (1996).
\bibitem{ioffe} B.L. Ioffe, Pisma ZhETF {\bf 42} (1985) 266;
{\bf 43} (1986) 316.
\bibitem{SVZ}  M.A. Shifman, A.I. Vainshtein and V.I. Zakharov, Nucl. Phys.
{\bf B147} (1979) 385, 448.
\bibitem{f2} V.M. Belyaev and B.L. Ioffe, Nucl.Phys. {\bf B310} (1988) 548.
\bibitem{g1g2} V.M. Belyaev and B.L. Ioffe, Int.J.Mod.Phys. {\bf 6A} (1991) 1533.
\bibitem{h1} B.L. Ioffe and A. Khodjamirian, Phys. Rev. {\bf D51}
(1995) 3380.
\bibitem{lightcone} I.I. Balitsky, V.M. Braun, and A.V. Kolesnichenko,
Sov.J.Nucl.Phys. {\bf 44} (1986) 1028;
Nucl.Phys. {\bf B312} (1989) 509.
\bibitem{bj} V.M. Belyaev and M.B. Johnson,  E-Print Archive: hep-ph/9605279
(1996).
\bibitem{cz} V.L. Chernyak and A.R. Zhitnitsky, Phys.Rep. {\bf 112} (1984) 173.
\bibitem{er} A.V. Efremov and A.V. Radyushkin, Phys.Lett. {\bf B94} (1980) 245.
\bibitem{bl} G.P. Lepage and S.J. Brodsky, Phys.Rev. {\bf D22} (1980) 2157.
\bibitem{vogt} M. Gluk, E. Reya, A. Vogt, Z.Phys. {\bf C53} (1992) 651.
\bibitem{bf} V.M. Braun and I.E. Filyanov, Z.Phys. {\bf C48} (1990) 239.
\bibitem{gorsky} A.S. Gorsky, Sov.J.Nucl.Phys. {\bf 41} (1985) 1008;
ibid. {\bf 45} (1987) 512.
\bibitem{nsvz} V.A. Novikov, M.A. Shifman, A.L. Vainshtein and
V.I. Zakharov, Nucl.Phys. {\bf B237} (1984) 525.
\bibitem{lepage}R.K. Ellis, W. Furmanski, 
R. Petronzio Nucl.Phys. {\bf B212} (1983) 29; 
R.L. Jaffe, Nucl.Phys. {\bf B229} (1983) 205;
I.I. Balitsky, Phys.Lett. {\bf B124} (1983) 230;
I.I. Balitsky, V.M. Braun, Nucl.Phys. {\bf B311} (1989) 541.
\bibitem{all} P. Ball, V.M. Braun, H.G. Dosch, Phys.Rev. {\bf D48}
(1993) 2110;
V.M. Belyaev, A. Khodjamirian, R. Ruckl,
 Z.Phys. {\bf C60} (1993) 349;
 A. Ali, V.M. Braun, H. Simma, Z.Phys. {\bf C63}
(1994) 43;
A. Ali, V.M. Braun, Phys.Lett. {\bf B359} (1995) 223;
A. Khodjamirian, R. Ruckl, Nucl.Instrum.Meth. {\bf A368} (1995) 28; 
V.M. Belyaev,  Z.Phys. {\bf C65} (1995) 93;
 G. Eilam, I. Halperin, R.R. Mendel, Phys.Lett.
{\bf B361} (1995) 137;
T.M. Aliev, D.A. Demir, E. Iltan, N.K. Pak, Phys.Rev. {\bf D54} 
(1996) 857;
P. Ball, V.M. Braun, 
 Phys.Rev. {\bf D54} (1996)2182;
 T.M. Aliev, A. Ozpineci, M. Savci, preprint  METU-PHYS-HEP-96-22,  E-Print Archive: hep-ph/9610256 ;
A. Khodjamirian, R. Ruckl, E-Print Archive: hep-ph/9610367; 
T.M. Aliev, A. Ozpineci, M. Savci, preprint METU-PHYS-HEP-96-35,  E-Print Archive: hep-ph/9612480. 
\bibitem{bbkr} V.M. Belyaev, V.M. Braun, A. Khodjamirian
and R. Ruckl, Phys.Rev. {\bf D51} (1995) 6177.



\end{thebibliography}
\end{document}